\input{aipcheck}

\documentclass[
    ,final            
  ]
  {aipproc}

\layoutstyle{8x11single}

\begin{document}

\title{Prompt optical observations of GRBs with "Pi of the Sky" system}


\author{M.~Sokolowski}{
  address={The Andrzej Soltan Institute for Nuclear Studies, Ho\.za~69, 00-681~Warsaw, Poland}
}

\author{M.~Cwiok}{
  address={Faculty of Physics, University of Warsaw, Ho\.za~69, 00-681~Warsaw, Poland}
}

\author{W.~Dominik}{
  address={Faculty of Physics, University of Warsaw, Ho\.za~69, 00-681~Warsaw, Poland}
}

\author{J.~Juchniewicz}{
  address={Space Research Center of the Polish Academy of Sciences, Bartycka~18A, 00-716~Warsaw, Poland}
}

\author{G.~Kasprowicz}{
  address={Institute of Electronic Systems, Nowowiejska~15/19, 00-665~Warsaw, Poland}
}

\author{A.~Majcher}{
  address={The Andrzej Soltan Institute for Nuclear Studies, Ho\.za~69, 00-681~Warsaw, Poland}
}

\author{A.~Majczyna}{
  address={The Andrzej Soltan Institute for Nuclear Studies, Ho\.za~69, 00-681~Warsaw, Poland}
}

\author{K.~Malek}{
  address={Center for Theoretical Physics, Polish Academy of Science, Al. Lotnikow 32/46, Warsaw, Poland}
}

\author{L.~Mankiewicz}{
  address={Center for Theoretical Physics, Polish Academy of Science, Al. Lotnikow 32/46, Warsaw, Poland}
}

\author{K.~Nawrocki}{
  address={The Andrzej Soltan Institute for Nuclear Studies, Ho\.za~69, 00-681~Warsaw, Poland}
}

\author{R.~Pietrzak}{
	address={Space Research Center of the Polish Academy of Sciences, Bartycka~18A, 00-716~Warsaw, Poland}
}

\author{L.W.~Piotrowski}{
  address={Faculty of Physics, University of Warsaw, Ho\.za~69, 00-681~Warsaw, Poland}
}

\author{D.~Rybka}{
  address={The Andrzej Soltan Institute for Nuclear Studies, Ho\.za~69, 00-681~Warsaw, Poland}
}

\author{J.~Uzycki}{
  address={The Andrzej Soltan Institute for Nuclear Studies, Ho\.za~69, 00-681~Warsaw, Poland}
}

\author{R.~Wawrzaszek}{
  address={Space Research Center of the Polish Academy of Sciences, Bartycka~18A, 00-716~Warsaw, Poland}
}

\author{G.~Wrochna}{
  address={The Andrzej Soltan Institute for Nuclear Studies, Ho\.za~69, 00-681~Warsaw, Poland}
}

\author{M.~Zaremba}{
  address={Faculty of Physics, Warsaw Univ. of Technology, Koszykowa~75, 00-662~Warsaw}
}

\author{A.F.~\.Zarnecki}{
  address={Faculty of Physics, University of Warsaw, Ho\.za~69, 00-681~Warsaw, Poland}
}

\begin{abstract}
The "Pi of the Sky" prototype apparatus observed prompt optical emission from
extremely bright GRB080319B since the very beginning of the gamma emission.
The burst occurred at redshift z=0.937 and set the record of optical luminosity reaching 5.3 mag.
The position of the burst was observed before, during and after the explosion by
several telescopes and unprecedented coverage of optical light curve
has been achieved. The combination of these unique optical data with simultaneous gamma-ray 
observations provides a powerful diagnostic tool for the physics of the GRB explosion
within seconds of its start.
The "Pi of the Sky" prototype, working since 2004 in Las Campanas Observatory in Chile,
consists of 2 cameras observing same 20$^\circ$x20$^\circ$ fields in the sky with
time resolution of 10 seconds. The prototype reacts to GCN alerts, but
it has also its own algorithm for identification of optical flashes.
The final system covering field of view of Swift or Fermi satellite
will consist of 2 arrays of 16 cameras installed in a distance of about 100 km.
The system is currently under construction.  It will be a powerful tool for early 
optical observations of GRBs, allowing for optical observation 
of GRBs  before, during and after the gamma emission.
With the on-line data analysis in real time, it will identify short
optical flashes autonomously and will be able to distribute this information 
among the community.
In this contribution the general idea of the final version of the experiment and the most interesting results 
from the prototype are presented.
\end{abstract}

\maketitle


\section{INTRODUCTION}
Observation of prompt optical emission accompanying gamma-ray bursts (GRBs)
can give deeper insight into mechanisms of the radiation and 
the central engines powering these extreme explosions.
Despite impressive advances over the roughly three decades since GRBs were  
discovered \cite{Klebesadel}, catching prompt optical signal 
 still remains a matter of fortune. Very large progress has been
achieved since the first optical counterparts of GRBs were observed 
 in 1997 thanks to Beppo-SAX satellite. In that time faint optical counterparts 
were observed many hours or even days after the gamma-ray explosion.
However, early observation of GRB990123 by the ROTSE robotic telescope~\cite{rotse_990123}
 has proven that prompt optical signal can also be very significant.
Hunting for prompt optical emission from GRBs triggered dynamic development
of robotic telescopes. After obtaining alert from the Gamma-Ray Burst
Coordinates Network (GCN) \cite{gcn}, they can reach the position of the burst within several
dozen of seconds and in many cases even faster. 
Large progress in the area has been achieved after Swift satellite \cite{swift} 
was launched in November 2004. The Burst Alert Telescope (BAT) detector \cite{swift_bat} onboard Swift satellite can
immediately provide position of the bursts with precision of 3 arcmin. The
UVOT telescope onboard Swift can perform follow-up optical observations very
quickly. However, in both cases of the UVOT and ground based robotic
telescopes the time delay between the burst and first optical observations
is inevitable. Typically UVOT starts observations 60 seconds after the
bursts, reaction time is sometimes shorter in case of robotic telescopes depending on 
the angular distance from the bursts. 
In spite of the fact that an enormous progress in fast optical follow-ups has
been achieved, the very first moment of the bursts is still rather a "dark~era"
 as far as optical observations are concerned. 
There were only few events for which the optical signal was observed during
the gamma-ray emission. The most famous one was GRB990123 observed by ROTSE \cite{rotse_990123},
the other one was the~GRB041219 observed by RAPTOR \cite{raptor_041219}. 
These two events were observed very early quite serendipitously, the first
one occurred in the field of view (FOV) of the ROTSE wide field camera, 
the later was triggered by the precursor and the alert was sent to GCN
network before the main burst has started. 
Apparently a new approach is needed to illuminate the "dark~era" of prompt optical
observations. The solution was postulated by late prof. B. Paczynski, who 
proposed to use wide field robotic telescopes for observations of early
optical emission from GRBs \cite{paczynski_robots}. This idea was
beautifully confirmed by recent observation of the naked-eye GRB080319B
performed by the prototype of the "Pi of the Sky" system. The final system,
which is currently under construction, will be much more powerful tool for 
early optical observations of GRBs.

\section{"PI OF THE SKY" SYSTEM}

\subsection{Final System Design}
\begin{figure}
	\includegraphics[width=6.333in,height=2.115in]{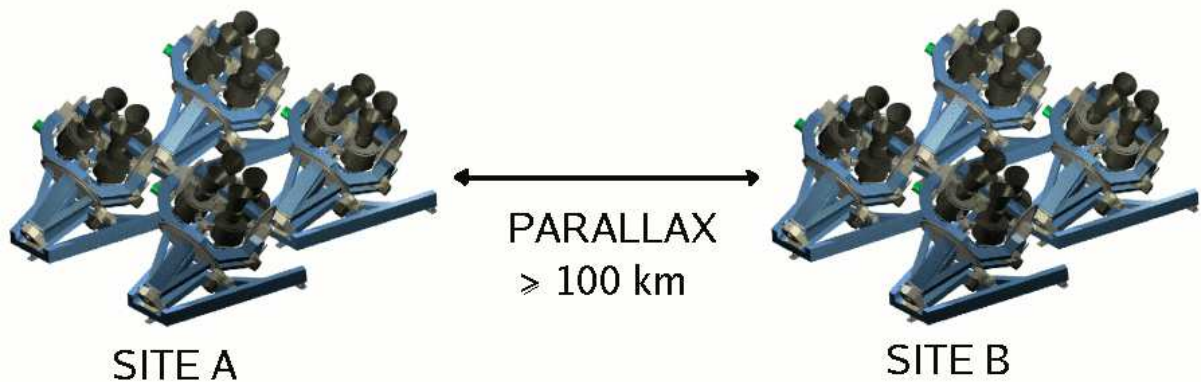}
  	\caption{Configuration of final system, able to cover whole Swift's field of view.}
	\label{fig_final_pi}
\end{figure}

The idea of continuous observations of the very wide field is to be realized by
 the "Pi of the Sky" system. The system will consist of two
farms of 16 cameras (Fig.\ref{fig_final_pi}), installed in a distance of $\ge$ 100 km. 
Each camera will cover field of view of 20$^\circ\times$20$^\circ$, 
which will result in a total sky coverage of $\sim$2 steradians.
This corresponds to field of view of Swift BAT \cite{swift_bat} and Fermi LAT \cite{fermi_lat}
detectors. Every GRB detected by the Swift, in the observing range of the "Pi~of~the~Sky"
system, will already be in its field of view. 
It is expected that $\sim$1/5 of all GRBs will occur in the FOV of the
system \cite{msok_phd}. They will be observed with 10 s time resolution since the very beginning of the burst, 
without any time delay normally needed to re-point the telescope. Moreover, the area of
interest will be observed even before the GRB, allowing to observe or
determine the upper limits for brightness of optical precursor.
The algorithms will analyze the images in the real time in search for short
optical flashes \cite{msok_phd}. 
The two sets of 16 cameras will observe the same part of the sky. The large 
distance between two sets is needed by the algorithms to use parallax to 
reject short optical flashes caused by artificial satellites and other near Earth sources.

The price for wide field observations is the limiting magnitude, which is
~12$^m$ on single 10 s exposure and reaches 14-15$^m$ on 20 averaged images.
However, predictions for the final system are very promising, the conservative
estimation gives $\sim$~2~GRB events per year to occur in its FOV 
and bright enough to be detected \cite{msok_phd}.
The number of positive detections can be even larger, because Fermi
satellite was not taken into account in this analysis. Another interesting perspective is a
possibility of observing optical counterparts of short GRBs. Recent
observations of optical counterparts of short GRBs indicate that they are
much dimmer then long ones, but they were never observed in the very first
moment of the explosion.
The system is currently under construction, it is planned to be installed in
2009 in Europe, probably somewhere in the Mediterranean region.

\subsection{Prototype working in Las Campanas Observatory}
Design and development of the prototype was the first step towards the final "Pi of the Sky" system.
The prototype was installed in Las Campanas Observatory in Chile in June 2004. 
It consists of two custom designed, low noise cameras \cite{kasprowicz} installed on a paralactic mount.
Each camera has a CCD with 2000 x 2000 pixels of 15$\times$15$\mu m^2$. 
They are equipped with CANON EF f = 85 mm, f/d = 1.2 photo lenses, 
giving pixel scale of 36 arcsec/pixel and covering 20$^\circ\times$20$^\circ$ field of view.
Currently the system observes sky in white light and only UV/IR-cut filter (transparent in 390-690 nm)
is used to suppress the background. The cameras continuously collect 10 s images
with 2 s breaks for readout. The values of limiting magnitude are the
same as expected for the final system (see above).
The cameras work in coincidence and observe the same field in the sky. 
The system works fully autonomously, it sends 
 SMS and e-mail to an operator in Warsaw, in case of GRB alert or any problems.  
For most of the time the telescope follows center of the Swift satellite's FOV. 
The coordinates of Swift's pointing are obtained from the GCN network. 
In case Swift's FOV cannot be observed, an alternative target like Integral or interesting object from list of 
blazars or AGNs \cite{gtn} is chosen. Twice a night whole sky scan is performed.
The on-line algorithm analyzes subsequent images in real time in search for short
optical flashes of cosmic origin. 

\subsection{Data Analysis}
The data analysis in the "Pi of the Sky" project consists of on-line and
off-line parts. The on-line algorithm analyzes images in real time in order to
find short optical flashes on the time scale of seconds or more. The algorithm was based on idea of multilevel
triggering system typical for high-energy physics experiments. 
Subsequent images are compared in order to find objects which appear in
the new image and were absent on series of previous images. 
After first two steps of the algorithm all pixels suspected of being new
objects in the sky are identified. Then several cuts are applied in order to
reject background candidates. The most important cuts are :

\begin{itemize}
\item \textbf{Coincidence} - this cut requires new object to be detected in
both cameras at the same position in the sky. It rejects mostly cosmic rays hitting one of CCD chips.
\item \textbf{SatCatalog}  - optical flashes due to artificial satellites are rejected by
using catalog of orbital elements retrieved daily from the Internet.
\item \textbf{StarCatalog} - rejects fluctuations of edges of bright stars according to star catalog
\item \textbf{Track} - fits tracks to objects from many subsequent images (typically 200) in
order to reject moving objects (i.e. satellites or planes) which could not
be rejected by previous \textbf{SatCatalog} criteria
\item \textbf{Shape} - Point Spread Function (PSF) of the event is tested and elongated objects are rejected.
\end{itemize}

The above cuts allow to suppress background efficiently without losing
too much signal. The algorithm was tested on real data. The efficiency
of optical flashes identification strongly depends on the weather conditions, reaching 80\% 
for flashes brighter than 9$^m$ during clear, moonless nights. The details of the algorithm can be found elsewhere \cite{msok_phd}.
This algorithm automatically identified about 200 short optical flashes 
 of unknown origin. Two events were unambiguously confirmed as flashes of astrophysical origin. 
The first one was an outburst of flare star CN Leo observed on 2.04.2005 \cite{aip_torun2005}.
The greatest success of the algorithm was identification of optical flash related to GRB080319B.
Most of the optical transients were not cross identified with any 
astrophysical events from other experiments. Observations of short optical
flashes my be very useful tool for identification of orphaned prompt optical
signal from GRBs. Current measurements from the prototype limit number of 
bright optical flashes related to GRB-like events to $\le$ 4 /day/4$\pi$ \cite{msok_phd}.

The raw data is reduced by custom designed reduction pipeline (\cite{msok_phd},\cite{asas})
 and measurements of star brightnesses are catalogued in a database. The
database, consisting of millions of stars brightness measurements, is analyzed by the
off-line algorithms (\cite{msok_phd},\cite{kasia_wilga}) in order to find 
new objects which appeared in the sky or brightening of existing
objects in time scales ranging from minutes to days. 
Several nova stars were recently discovered by this kind of analysis \cite{pi_web_page}.

\section{PROMPT OPTICAL OBSERVATIONS OF GRBs}

\subsection{Observation of the GRB080319B}
\begin{figure}
	\includegraphics[width=2.7in,height=2.7in]{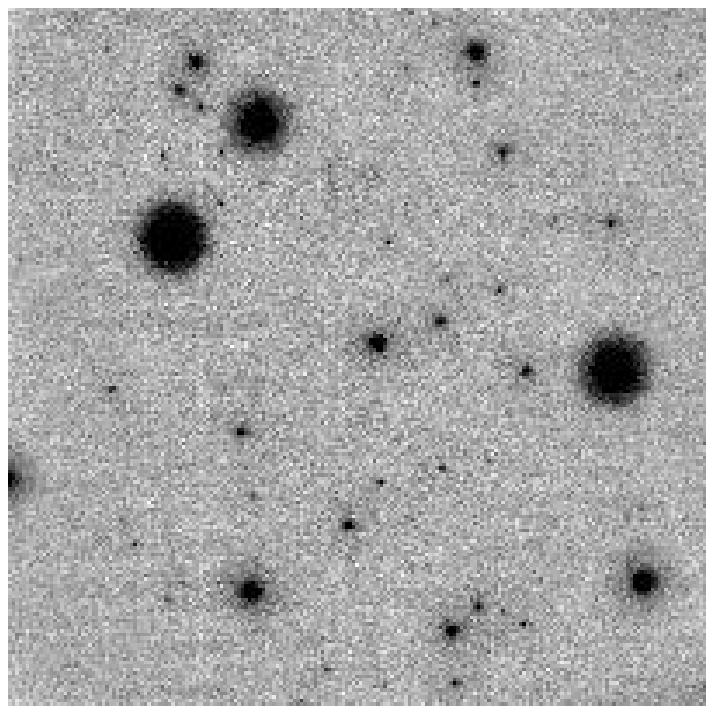}
	\includegraphics[width=2.7in,height=2.7in]{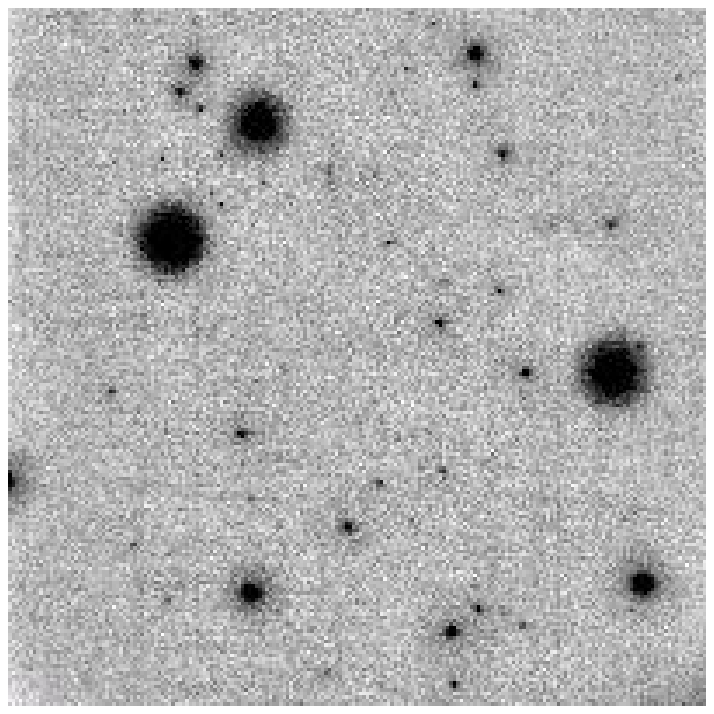}
  \caption{Optical counterpart of the GRB 080319B, left image with very bright optical signal in the center (taken at 06:14:03 UT)
and right image without optical transient (taken 10 minutes after the burst)}
	\label{fig_grb080319B_ot}
\end{figure}

\begin{figure}

	\includegraphics[width=3.55in,height=2.4in]{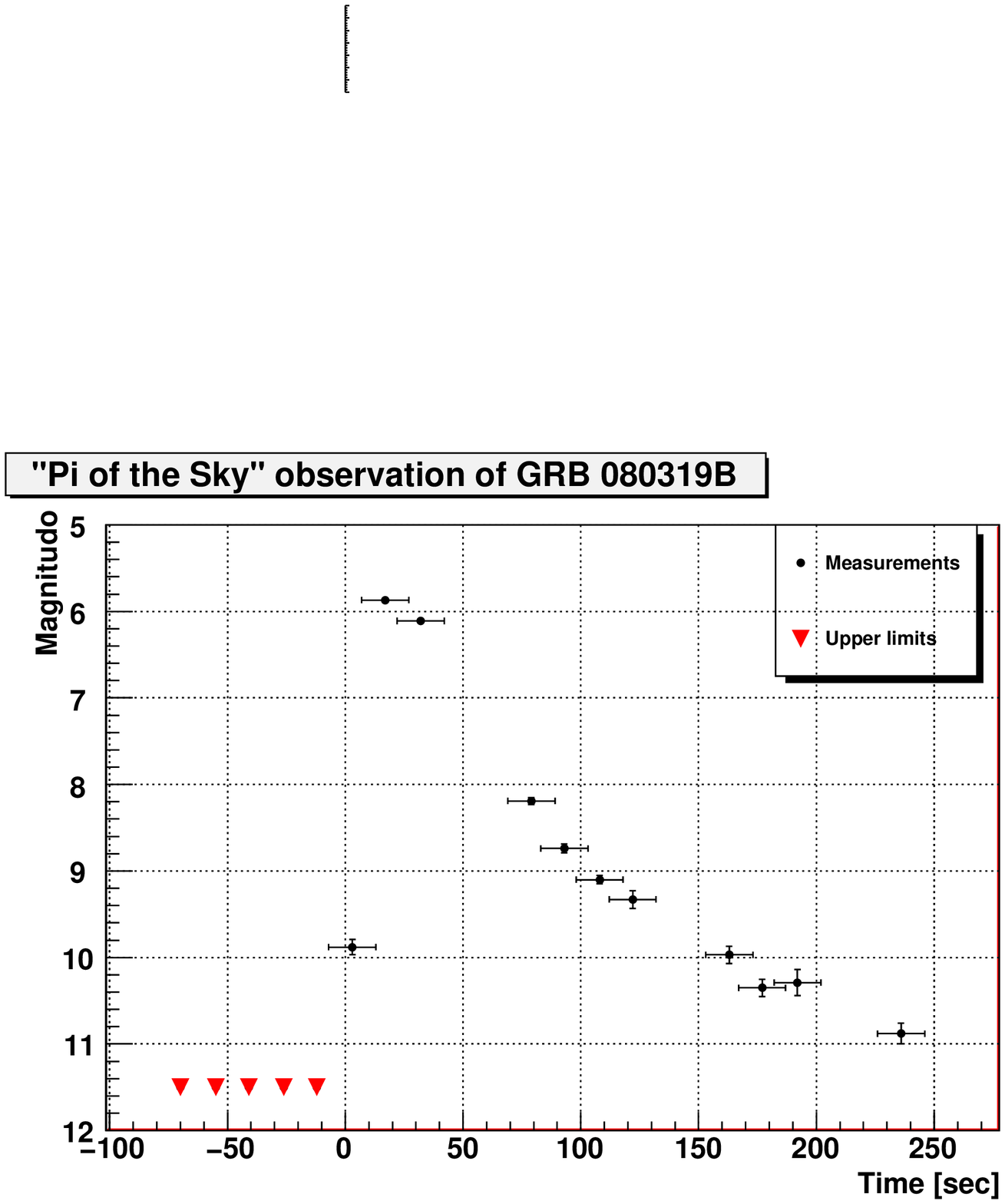}
	\includegraphics[width=3.55in,height=2.4in]{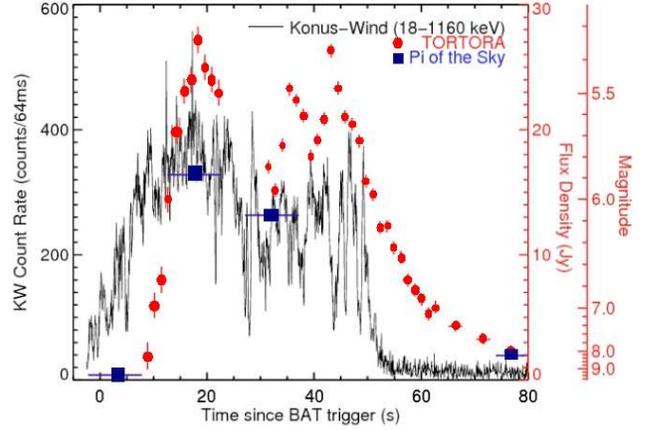}
  \caption{Optical lightcurve of GRB 080319B measured by the "Pi of the Sky"
prototype (left plot, points are centered in central time of the exposure and 10 s error
bar corresponds to exposure time) and early optical measurements with
$\gamma$-ray lightcurve superimposed (right plot from \cite{grb080319b_astroph})}
\label{fig_grb080319B_lc}
\end{figure}

On March 19, 2008, at 06:12:49 Universal Time the BAT detector onboard the Swift satellite \cite{swift_bat} was triggered by an
intense pulse of gamma rays from GRB 080319B \cite{racusin_gcn}. The BAT
localization distributed over the GCN at 06:13:06 was received by the "Pi of
the Sky" system, while the follow-up of previous alert (GRB 080319A) was
still in progress. The angular distance of GRB 080319B from position of GRB
080319A was $\alpha \approx$11.76$^\circ$, which was small enough to catch
GRB 080319B in the FOV of the telescope.
The "Pi of the Sky" telescope has observed the area of
interest for 23 min 33 sec before the burst. The first exposure on which 
optical signal from GRB 080319B was detected has started at 06:12:47
Universal Time, two seconds before the outburst was detected by the BAT. 
Very bright optical flash has been automatically detected by the on-line
algorithm searching for short optical transients.
The optical signal was bright enough to be observed by the "Pi of the Sky"
telescope for next 4 minutes (Fig. \ref{fig_grb080319B_ot}), after that time it faded below the limiting
magnitude of the system \cite{pi_gcn_grb080319B}. The maximal brightness of
the optical transient reached 5.3$^m$ which means that the event which
occurred in a distance of almost 7.5 billion years could be visible with a 
naked-eye.
The optical flux was normalized to V~filter magnitude of
the neighboring stars using the procedure described in \cite{msok_phd}.
The measurements are listed in Table \ref{tab_grb080319B_lc}, 
systematic errors shown in this table are due to normalization of unfiltered
observations to V~filter brightness.
The optical lightcurve is shown in Figure \ref{fig_grb080319B_lc}. 
Observation of GRB 080319B confirms strongly that the wide field
 cameras are the best and probably the only effective way for 
 observations of prompt optical signal. The unprecedented coverage of the
optical lighcurve was a key for understanding the physics of the event. 
The results of multiwavelength analysis of GRB 080319B show that early optical observations can have
 very significant impact on modeling the phenomena \cite{grb080319b_nature}.
As it can be seen from Fig. \ref{fig_grb080319B_lc} optical signal starts and ends in the
same moment as gamma emission. This can prove that optical signal is
produced in the same region as gamma rays. On the other hand optical flux is
3-4 orders of magnitude larger than gamma ray spectrum extrapolated to
optical region, suggesting different production mechanisms.
Even observation of the one single GRB can have large influence
on the existing models of these mysterious events \cite{grb080319b_nature}.
This exhibits how much the GRB understanding suffers from lack of the early optical data.

\begin{table}[htbp]
\begin{tabular}{ccccccc}
\hline
  \tablehead{1}{c}{b}{Start Time (UT)}
  & \tablehead{1}{c}{b}{V}
  & \tablehead{1}{c}{b}{\boldmath{$\delta_{stat}$} V}
  & \tablehead{1}{c}{b}{\boldmath{$\delta_{sys}$} V}
  & \tablehead{1}{c}{b}{flux [Jy]}
  & \tablehead{1}{c}{b}{\boldmath{$\delta_{stat}$} f [Jy]}
  & \tablehead{1}{c}{b}{\boldmath{$\delta_{sys}$} f [Jy]}   \\
\hline
 $\le$ 06:12:32 & > 11.5 & - & - &  <0.095 & - & - \\
\hline
 06:12:47 & 9.88 & 0.09 & 0.14 & 0.42 & 0.04 & 0.06 \\
\hline 
 06:13:01 & 5.87 & 0.03 & 0.27 & 17.0 & 0.5 & 4.24  \\
\hline
 06:13:16 & 6.11 & 0.03 & 0.27 & 13.6 & 0.4  & 3.38 \\
\hline
 06:14:03 & 8.19 & 0.04 & 0.17 & 2.00 & 0.07 & 0.31 \\
\hline
 06:14:17 & 8.74 & 0.05 & 0.16 & 1.20 & 0.06 & 0.18 \\
\hline
 06:14:32 & 9.10 & 0.05 & 0.15 & 0.87 & 0.04 & 0.12 \\
\hline
 06:14:46 & 9.33 & 0.10 & 0.14 & 0.70 & 0.06 & 0.09 \\
\hline
 06:15:27 & 9.97 & 0.10 & 0.14 & 0.39 & 0.03 & 0.05 \\
\hline
 06:15:41 & 10.35 & 0.10 & 0.15 & 0.27 & 0.03 & 0.04 \\
\hline
 06:15:56 & 10.29 & 0.15 & 0.15 & 0.29 & 0.04 & 0.04 \\
\hline
 06:16:40 & 10.88 & 0.12 & 0.15 & 0.17 & 0.02 & 0.02 \\
\hline
\end{tabular}
\caption{Brightness of optical counterpart of GRB 080319B on 10 s images}
\label{tab_grb080319B_lc}
\end{table}

\subsection{Early limits on other GRBs}
The "Pi of the Sky" system is working since June 2004. Since then until now (2008.12.18) about 360 
GRBs with known position were observed by the satellites, most of them occurred outside the range of the telescope and could not be
observed. In three cases (except GRB 080319B) the burst occurred in the FOV of
the telescope and upper limits for prompt optical emission have been
determined (Tab. \ref{tab_grb_limits}). In some other cases GRB position
was reached in less than two minutes, such limits are also listed in the Table \ref{tab_grb_limits}.
In none of these cases optical counterpart was observed. 

\begin{table}[htbp]
\begin{small}
\begin{tabular}{cccccc}
\hline
  \tablehead{1}{c}{b}{GRB}
  & \tablehead{1}{c}{b}{Reaction Time [seconds]}
  & \tablehead{1}{c}{b}{Before}
  & \tablehead{1}{c}{b}{During}
  & \tablehead{1}{c}{b}{After}
  & \tablehead{1}{c}{b}{During or After [Jy]} \\
\hline
 \textbf{040825A} & <0 & 10.0$^m$ & 12.0$^m$ &  9.5$^m$ & 0.06 \\
\hline
 \textbf{050412}  & <0 & 11.5$^m$ & 11.0$^m$ & 11.5$^m$ & 0.15 \\
\hline
 \textbf{070521}  & <0 & 12.2$^m$ & 12.6$^m$ & 12.5$^m$ & 0.035 \\
\hline
 080916A & 66  & - & - & 11.3 & 0.11 \\
\hline
 080805  & 83  & - & - & 12.2 & 0.05 \\
\hline
 080804  & 101 & - & - & 12.0 & 0.06 \\
\hline
 080409  & 96  & - & - & 12.0 & 0.06 \\
\hline
 080310  & 99  & - & - & 12.4 & 0.04 \\
\hline
 081007  & 71  & - & - & 12.1 & 0.05 \\
\hline
 070913  & 110 & - & - & 12.6 & 0.034 \\
\hline
 060719  & 65  & - & - & 12.8 & 0.03 \\
\hline
 050607  & 60  & - & - & 12.5 & 0.04 \\
\hline
 050522  & 75  & - & - & 11.0 & 0.15 \\ 
\hline
\end{tabular}
\end{small}
\caption{Upper limits on brightness of prompt emission for GRBs which occurred in FOV of the "Pi of the Sky" system
or have been observed less then 2 minutes after the GRB}
\label{tab_grb_limits}
\end{table}

\section{SUMMARY}
The prototype of the "Pi of the Sky" system working in Las Campanas
Observatory in Chile has observed prompt optical emission from GRB 080319B. 
Unprecedented coverage of the optical lightcurve allowed for precise analysis of this event.
It turns out that even single, well measured burst can challenge the
mainstream models of GRBs.  
This observation have proven that the best way for observing early optical signal from GRBs is to use wide field cameras.
The final "Pi of the Sky" system able to cover whole FOV of the Swift satellite is
currently under construction. It is expected that $\sim$1/5 of all Swift's GRBs
will occur in the FOV of the system and at least 2 long GRBs per year will be bright
enough to be observed by the system.
Another great opportunity of such farm of wide field cameras is a possibility of observing
short GRBs in the very first moment of the explosion.


\begin{theacknowledgments}
We are very grateful to G.~Pojmanski for access to ASAS dome and sharing his experience with us.
We would like to thank the staff of the Las Campanas Observatory for their help during the
 installation of the apparatus.
This work was financed by the Polish Ministry of Science in 2005-2008 as a
research project.

\end{theacknowledgments}



\bibliographystyle{aipproc}   

\bibliography{pi_of_the_sky}

\IfFileExists{\jobname.bbl}{}
 {\typeout{}
  \typeout{******************************************}
  \typeout{** Please run "bibtex \jobname" to optain}
  \typeout{** the bibliography and then re-run LaTeX}
  \typeout{** twice to fix the references!}
  \typeout{******************************************}
  \typeout{}
 }

\end{document}